\documentclass[aps,prb,twocolumn,superscriptaddress,showpacs]{revtex4}

\usepackage{graphicx}% Include figure files

\begin{document}
%\draft command makes pacs numbers print
%\draft
\title{Valence band orbital polarization in III-V ferromagnetic semiconductors}
\author{A.~A.~Freeman}
\affiliation{School of Physics and Astronomy, University of Nottingham, Nottingham NG7 2RD, United Kingdom}
\author{K.~W.~Edmonds}
\affiliation{School of Physics and Astronomy, University of Nottingham, Nottingham NG7 2RD, United Kingdom}
\author{G.~van~der~Laan}
\affiliation{STFC Daresbury Laboratory, Warrington WA4 4AD, United Kingdom}
\affiliation{Diamond Light Source, Didcot OX11 0DE, United Kingdom}
\author{R.~P.~Campion}
\affiliation{School of Physics and Astronomy, University of Nottingham, Nottingham NG7 2RD, United Kingdom}
\author{N.~R.~S.~Farley}
\affiliation{School of Physics and Astronomy, University of Nottingham, Nottingham NG7 2RD, United Kingdom}
\author{A.~W.~Rushforth}
\affiliation{School of Physics and Astronomy, University of Nottingham, Nottingham NG7 2RD, United Kingdom}
\author{T.~K.~Johal}
\affiliation{STFC Daresbury Laboratory, Warrington WA4 4AD, United Kingdom}
\author{C.~T.~Foxon}
\affiliation{School of Physics and Astronomy, University of Nottingham, Nottingham NG7 2RD, United Kingdom}
\author{B.~L.~Gallagher}
\affiliation{School of Physics and Astronomy, University of Nottingham, Nottingham NG7 2RD, United Kingdom}
\author{A.~Rogalev}
\affiliation{European Synchrotron Radiation Facility, Boite Postale 220, F-38043 Grenoble Cedex 9, France}
\author{F.~Wilhelm}
\affiliation{European Synchrotron Radiation Facility, Boite Postale 220, F-38043 Grenoble Cedex 9, France}

\date{\today}%

\begin{abstract}
The element-specific technique of x-ray magnetic circular dichroism (XMCD) is used to directly determine the
magnitude and character of the valence band orbital magnetic moments in (III,Mn)As ferromagnetic semiconductors.
A distinct dichroism is observed at the As $K$ absorption edge, yielding an As $4p$ orbital magnetic moment of
around $-$0.1 $\mu_B$ per valence band hole, which is strongly influenced by strain, indicating its crucial influence on the magnetic anisotropy. The dichroism at the Ga $K$ edge
is much weaker. The $K$ edge XMCD signals for Mn and As both have positive sign, which indicates the important
contribution of Mn $4p$ states to the Mn $K$ edge spectra.
\end{abstract}

\maketitle

Magnetocrystalline anisotropy (MCA) plays a central role in technological applications of ferromagnetism, from
permanent magnetic materials to ultrathin films. Since the MCA influences properties such as domain wall width, spin transfer torque and magnetization dynamics, it is crucial for determining the
spintronic functionalities of a material. The MCA has at its heart the spin-orbit interaction leading to a coupling between the magnetization and the crystal lattice. The MCA energy is directly related to the anisotropic part of the spin-orbit interaction \cite{vdl1}, and is shown to be proportional to the
difference in the orbital magnetic moment $m_{\mathrm{orb}}$ along the easy and hard magnetic axes when one of the spin
sub-bands is filled \cite{bruno}. In cubic transition metal ferromagnets, the
MCA is typically rather small (corresponding to a few $\mu$eV per atom \cite{daalderop}), but may be
substantially enhanced when the symmetry of the system is reduced \cite{gambardella}. Several new insights into the origin of MCA in a range of materials have been delivered by x-ray magnetic circular and linear dichroism (XMCD and XMLD), due to their element-specific sensitivity to orbital moments \cite{thole} and spin-orbit coupling anisotropies \cite{stohr,durr}, respectively.

A particularly interesting MCA is found in diluted magnetic semiconductors such as Ga$_{1-x}$Mn$_x$As, where a
carrier-mediated interaction between Mn acceptors leads to ferromagnetic order, with a Curie temperature
$T_C$ up to 173~K for $x$$\approx $$7$\%. At low concentrations,
Mn acceptors in GaAs possess a $d^5$ electronic configuration, with negligible single-ion anisotropy associated
with the Mn core \cite{almeleh}. However, in the high concentration ferromagnetic regime, a large uniaxial MCA
is observed, which is strongly influenced by epitaxial strain \cite{shen}. The MCA in (Ga,Mn)As
has been explained semiquantitatively, by considering the combined effects of strain and exchange splitting on a
model valence band built from $s,p$ orbitals of the host ions \cite{abolfath}. Within this model, the
MCA (and related anisotropic magnetoresistance effects \cite{rushforth}) results from the induced anisotropy of
the strongly spin-orbit coupled ($j$=3/2) valence hole states, with predominantly As $4p$ character. The model also predicts a sizeable magnetization of the hole subsystem, of opposite sign to the Mn magnetization \cite{sliwa}.

Ab-initio calculations generally agree that the ferromagnetism in (Ga,Mn)As is
mediated by an antiferromagnetic $p$-$d$ exchange coupling, but also point to a significant $d$ character of the
valence band holes \cite{sandratskii,wierzobowska,schulthess}. The amount of $d$ weight at the Fermi energy depends
strongly on the details of the calculations, and different studies yield values of the $p$-$d$ coupling which
may differ by as much as a factor of two. Also, due to computational effort required, ab-initio
calculations generally do not include spin-orbit effects and thus are unable to account for the observed
magnetic anisotropy. Multiband tight-binding calculations incorporating spin-orbit interaction
suggest that hybridization between Mn $3d$ and GaAs valence bands results in resonances with strong orbital
polarization close to the valence band edge \cite{tang}.

The antiferromagnetic coupling between Mn local moments and holes has been reported in optical magnetic circular
dichroism studies \cite{szczytko}. However, since such measurements involve interband transitions, a direct element-specific and shell-specific determination of the
states involved in the transitions is not possible. More recently, As and Ga $L_{2,3}$ edge XMCD studies clearly
demonstrated the presence of induced magnetic moments on the host ions \cite{keavney}. Here the
excitation is from the $2p$ core level of either As or Ga, which are separated in energy by $\approx$~230~eV,
allowing an element-specific assignment of the measured signals: it was found that the As signal is larger than
the Ga signal by a factor of $\sim$7, and opposite in sign. However, due to the electric-dipole selection
rules governing the transition (specifically, $\Delta l = \pm 1$) the $4p$ valence holes, which are the principal states governing the spintronic functionalities of the material, are not probed in $L$-edge absorption.

Here, we exploit $K$-edge XMCD to directly probe the magnetic anisotropy of the $4p$ valence states in (III,Mn)As ferromagnetic semiconductors. Since $K$-edge absorption involves
excitations from the $1s$ ($l$=0) core level, a significant XMCD will occur only if there is an orbital
polarization of the probed valence states \cite{igarashi}. We present the first direct determination of
the As and Ga $4p$ orbital magnetic moments in III-V ferromagnetic semiconductors.

The penetration depth of x-rays at the Ga and As $K$-edges is several tens of microns, which is much larger than
the thickness of typical (Ga,Mn)As films. Therefore, Ga and As absorption signals from the substrate must be
avoided. We investigated two different samples: (a) 1~$\mu$m thick (Ga,Mn)As film grown on an AlAs buffer
layer on GaAs(001), released from the substrate by etching the AlAs layer, and re-mounted on sapphire; (b)
0.5~$\mu$m thick (In,Ga,Mn)As film on InP(001), which is nearly lattice-matched to the substrate by varying the
In content in the film \cite{slupinski}. X-ray diffraction measurements show that the latter film is under small
compressive strain. The films were grown by low-temperature (200-250$^\circ$C) molecular beam epitaxy,
and the nominal Mn concentration (estimated from the Mn flux during growth) was $\sim$8\%. SQUID magnetometry
measurements show that the (Ga,Mn)As and (In,Ga,Mn)As films are ferromagnetic with $T_C$ of 54 K and 25 K, respectively.

The XMCD measurements were performed on ID12 of the European Synchrotron Radiation Facility (ESRF) at Grenoble.
Ga, As and Mn $K$-edge absorption spectra were obtained from total fluorescence yield measurements using 98\%
circularly polarized x-rays. The XMCD signal was obtained from the difference in absorption for parallel and
antiparallel alignments of the x-ray polarization vector with respect to an external magnetic field.
Measurements, taken over several hours, were averaged in order to improve the signal-to-noise ratio. To avoid experimental
artefacts, the external magnetic field direction and the x-ray helicity were alternately flipped. The
measurements were performed at a sample temperature of 10~K, under a magnetic field of $\pm$2~T along the beam, which was
either perpendicular or at grazing incidence (15$^\circ$) to the sample surface. Self-absorption effects are accounted for in the quantitative analysis, and are very small as the probing and escape depths are much larger than the film thickness.

Figure 1(a-c) shows the As $K$-edge absorption spectrum from the (Ga,Mn)As film, and the Ga and As $K$-edge
absorption spectra from the (In,Ga,Mn)As film. A clear dichroism is observed at the onset of the As
absorption edge, indicating a polarization of the As $4p$ states at the valence band edge. The position and
shape of the XMCD spectrum is similar for the two films, and also qualitatively similar to the main feature
observed in As $L_3$ XMCD from (Ga,Mn)As \cite{keavney}. From the positive sign of the dichroism we are able to
conclude that the As $4p$ orbital polarization is antiparallel to the net magnetization. We also observe a
substantial anisotropy of the As XMCD for the strained (In,Ga,Mn)As film, with a larger signal for grazing
incidence than for normal incidence. At the Ga edge the XMCD is much weaker, and is scarcely visible above the
noise level.

A magneto-optical sum rule relates the integrated XMCD signals across an absorption edge to element-specific
orbital magnetic moments per atom \cite{thole}. However, a quantitative application of this rule requires an
accurate knowledge of the number of occupied $4p$ states, $n_{4p}$, as well as a separation of the $1s$ to $4p$
and $1s$ to continuum contributions to the absorption edge. Here we take $n_{4p}$ to be equal to 3, and describe
the continuum background by a step function positioned under the highest intensity point of the absorption
spectrum. Note that since the sum rule analysis is concerned only with integrated intensities, the precise shape
of the background function is unimportant. Since the estimated 10-20\% uncertainties introduced by the above
assumptions are systematic, the relative moments obtained from the different samples and orientations can be
compared with good accuracy.

Results of the sum rule analysis are given in Table I. The obtained As $4p$ orbital moments are around
10$^{-3}$~$\mu_B$ per As atom, while the Ga $4p$ orbital moment is an order of magnitude lower. From room
temperature Hall effect measurements, we estimate that the hole concentration in the films is around
10$^{20}$~cm$^{-3}$, in agreement with the values obtained at low temperatures and high magnetic fields for (Ga,Mn)As films with similar Mn doping \cite{edmonds2}. Since the concentration of As is around 2x10$^{22}$~cm$^{-3}$, this gives an As  $4p$ orbital moment per valence band hole of around $-$0.1 to $-$0.2~$\mu_B$.

For the (In,Ga,Mn)As film, the As $4p$ orbital moment at normal incidence is only $\sim$60\% of the value
obtained at grazing incidence. Since XMCD measures the projection of the orbital moment along the photon
wavevector, the measured anisotropy indicates that, under a saturating magnetic field, the As $4p$ magnetization
is larger for in-plane directions than for perpendicular to the plane.  Model calculations have shown that in the absence of strain, the crystalline anisotropy of the hole magnetization is negligible \cite{sliwa}. Therefore, the observed anisotropy of the As $4p$ magnetization is ascribed to the small compressive strain. This leads to an in-plane easy magnetic axis, as confirmed by magnetometry and magnetotransport measurements. The magnetic easy axis is therefore aligned with the direction where the As $4p$ orbital moment is largest. In constrast to the As $4p$ moment, angle-dependent XMCD measurements at the Mn $L_{2,3}$ edges of (Ga,Mn)As yielded no discernable anisotropy of the small Mn $3d$ orbital moment within the experimental uncertainty
\cite{edmonds}. This indicates that it is the anisotropy of valence holes with As $4p$ character in (III,Mn)As
ferromagnetic semiconductors that gives the dominant contribution to the MCA, in qualitative agreement with
hole-fluid models of ferromagnetism in these systems \cite{abolfath, dietl}.

Figure 2 shows the Mn $K$-edge x-ray absorption and XMCD spectra from the (Ga,Mn)As and (In,Ga,Mn)As films. The prominant
part of the Mn XMCD consists of a double peak structure with a splitting of $\sim$1.4~eV, centered around the
pre-edge feature in the x-ray absorption spectrum. The Mn and As $K$-edge XMCD signals have the same sign, and are
of comparable magnitude.

Pre-edge structures are a typical feature of $K$-edge absorption spectra from systems with open $3d$ shells, and
are usually associated with $1s \to 3d$ transitions. While being formally electric-quadrupole transitions that
are usually very weak, a stronger electric-dipole allowed character can mix in due to indirect hybridization
between $3d$ and $4p$ orbitals on adjacent sites through the ligand states \cite{bridges} or due to direct
$p$-$d$ hybridization on the excited atom. The latter mechanism may occur only in non-centrosymmetric
environments \cite{westre}, such as the $T_d$ symmetry of the Mn site in (Ga,Mn)As and (In,Ga,Mn)As. Mn
$L_{2,3}$ XMCD measurements have shown that the orbital and spin moments of the Mn $3d$ shell are parallel
\cite{edmonds,ueda}. Therefore, the sign of the Mn $K$-edge XMCD is interpreted as being due to on-site mixing
between Mn $3d$ and $4p$ states, where the Mn $3d$ and $4p$ orbital moments are antiparallel. Furthermore, since
the Mn $4p$ shell is much less than half-filled, Hund's third rule implies that orbital and spin moments are
antiparallel so that the Mn $3d$ and $4p$ spin moments are parallel.

Density-functional calculations of Ga, Mn, and As $K$-edge XMCD from (Ga,Mn)As have previously been reported
\cite{wu}. While the magnitudes of the calculated XMCD for As and Mn are in reasonable agreement with our
experimental results, the details of the spectra are rather different. In particular, the calculated Mn XMCD
showed a sharp negative peak at the onset of the pre-edge \cite{wu}, which is absent in the experimental
spectrum shown in Fig. 2. Such a feature is however observed in the XMCD spectrum of substitutional Mn in the wider bandgap
material GaN \cite{sarigiannidou}. This suggests that the calculations of Ref.~\onlinecite{wu} have
overestimated the $p$-$d$ hybridization, giving a spuriously large Mn 3$d$ component to the valence band hole
\cite{wierzobowska}.

Comparing the present $K$-edge results to previous $L_{2,3}$-edge XMCD measurements of (Ga,Mn)As, we can draw
some conclusions about the alignment of the element- and shell-resolved spin and orbital magnetic moments. These conclusions are summarized in Table II. As discussed above, the present results show that the As and Ga $4p$ orbital moments are antiparallel to the Mn $3d$ spin moment, and imply that the Mn $4p$ orbital and spin moments are respectively antiparallel and parallel to the Mn $3d$ spin moment. In addition, the Fermi energy is significantly lower than the spin-orbit splitting of the valence band in the present samples, so that the valence holes have mostly $j$=3/2 character, with
parallel orbital and spin moments of the Ga and As $4p$ shells. Elsewhere, Ga and As $L_{2,3}$ edge measurements
indicated that their $4s$ spin moments are respectively parallel and antiparallel to the Mn $3d$ spin moment
\cite{keavney}. However, this interpretation neglects possible $2p$ to $3d$ transitions, which would give a
contribution of opposite sign. Finally, Mn $L_{2,3}$ measurements have shown that the Mn $3d$ spin and orbital
moments are parallel \cite{edmonds,ueda}.

In summary, we have used x-ray magnetic circular dichroism to obtain the sign and magnitudes of the element- and
shell-specific orbital magnetic moments in the III-V ferromagnetic semiconductors (Ga,Mn)As and (In,Ga,Mn)As.
The large As $4p$ orbital polarization is shown to give the important contribution to the strain-induced magnetic
anisotropy in (III,Mn)As ferromagnetic semiconductors. These results will form a useful basis for the development of
microscopic theories of spin-orbit coupling effects in ferromagnetic semiconductors and related systems. Technologically
important phenomena in these materials rely on the spin-polarization and the magnetic anisotropy, and the modelling of
these will ultimately need to conform to the correct magnetic coupling revealed by the present study.

\begin{table}
\begin{tabular}{llcc}
\hline \hline
Magnetic & Direction & Sample (a) &  Sample (b) \\
moment & & (Ga,Mn)As &  (In,Ga,Mn)As \\
\hline
 $m_{\mathrm{orb}}^{\mathrm{As}}$ & normal & $-$1.3$\pm$0.2  & $-$0.6$\pm$0.1 \\
$m_{\mathrm{orb}}^{\mathrm{Ga}}$ & normal & - & $-$0.08$\pm$0.04 \\
$m_{\mathrm{orb}}^{\mathrm{As}}$ & grazing & - & $-$1.0$\pm$0.2 \\
\hline \hline
\end{tabular}
\caption{As $4p$ and Ga $4p$ orbital magnetic moments ($\times$10$^{-3}$~$\mu_B$ per Ga or As atom) obtained
from sum rule analysis.} \label{table1}
\end{table}

\begin{table}
\begin{tabular} {lcccccc}
\hline \hline & Mn $3d$ &
 Mn $4p$ &
 As $4s^\star$ &
 As $4p$ &
 Ga $4s^\star$ &
 Ga $4p$ \\
 \hline
$m_{\mathrm{spin}}$ & $\uparrow$ & $\uparrow$ & $\downarrow$ & $\downarrow$ & $\uparrow$  &
 $\downarrow$ \\
$m_{\mathrm{orb}}$ & $\uparrow$ & $\downarrow$ & 0 &
 $\downarrow$ &
0 &
 $\downarrow$ \\
\hline \hline
\end{tabular}
\caption{Alignment of element- and shell-specific spin and orbital magnetic moments in (Ga,Mn)As, as determined
from XMCD measurements. Note that the orientations of the As and Ga $4s$ moments ($^\star$) are obtained from
$L_{2,3}$ edge measurements (Ref.~19), which (due to the dipole selection rule $\Delta l = \pm
1$) could alternatively be interpreted as being due to As and Ga $3d$ spin moments.}
\end{table}

\begin{figure}       % FIGURE 1
\includegraphics[trim = 10mm 20mm 20mm 20mm,clip, width=80mm, angle=0]{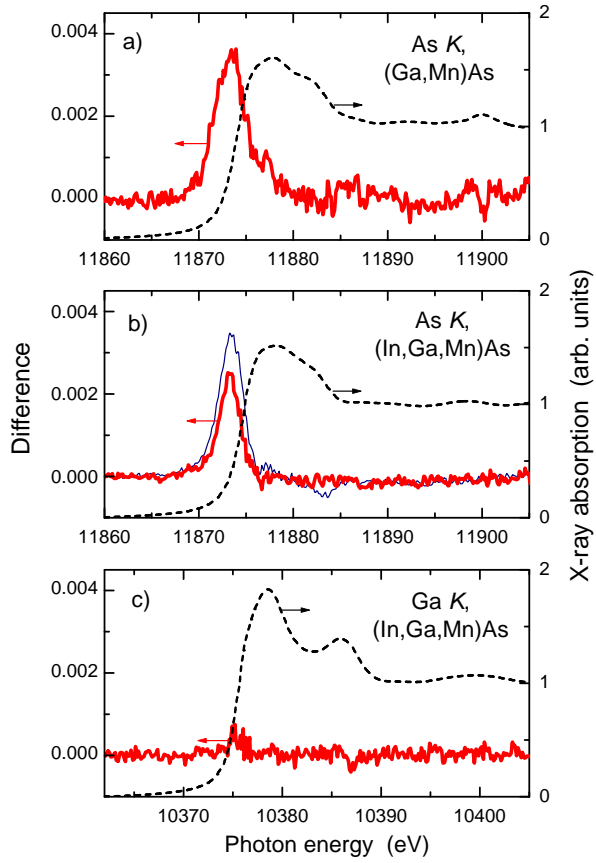}
\label{Fig. 1} \caption {(Color online) (a) XMCD spectra (drawn lines, left axes) and x-ray absorption spectra
(dashed lines, right axes) for (a) As $K$ edge of (Ga,Mn)As; (b) As $K$ edge of (In,Ga,Mn)As; (c) Ga $K$ edge of
(In,Ga,Mn)As. The thick red (gray) lines give the XMCD for normal incidence, and the thin blue (black) line in
(b) gives the XMCD for grazing incidence.}
\end{figure}

\begin{figure}         % FIGURE 2
\includegraphics[trim = 2mm 5mm 2mm 5mm,clip, width=80mm, angle=0]{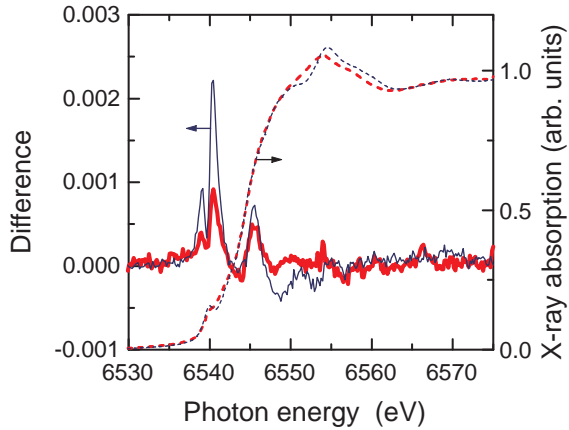}
\label{Fig. 2} \caption {(Color online) Mn $K$ edge XMCD spectra (drawn lines, left axis) and x-ray absorption
spectra (dashed lines, right axis) for (Ga,Mn)As (thin blue (black) lines) and (In,Ga,Mn)As (thick red (gray)
lines).}
\end{figure}

\end{document}